\documentclass[11pt]{article}

\usepackage[margin=1in]{geometry}
\usepackage{amsmath,amssymb}
\usepackage[nopatch]{microtype}
\usepackage{booktabs}
\usepackage{enumitem}
\usepackage{caption}
\usepackage{subcaption}
\usepackage{multirow,multicol}
\usepackage[ruled,noend]{algorithm2e}
\usepackage{graphicx}
\usepackage{hyperref}
\usepackage{natbib} 
\usepackage{float}

\title{Identifying Subgroup and Context Effects in Conjoint Experiments}

\author{
  Steven Wang$^{1}$ \and
  Isys Johnson$^{2}$ \and
  Jessica Grogan$^{2}$ \and
  Lalit Jain$^{3}$ \and
  Atri Rudra$^{2}$ \and
  Kyle Hunt$^{1}$ \and
  Kenneth Joseph$^{2}$\thanks{Corresponding author: \texttt{kjoseph@buffalo.edu}}
}

\date{\vspace{-0.5em}%
$^{1}$School of Management, University at Buffalo, Buffalo, NY, USA\\
$^{2}$Computer Science and Engineering, University at Buffalo, Buffalo, NY, USA\\
$^{3}$Foster School of Business, University of Washington, Seattle, WA, USA\\[0.5em]
\vspace{0.5em}
}

\begin{document}
\maketitle

\begin{abstract}
Conjoint experiments have become central to survey research in political science and related fields because they allow researchers to study preferences across multiple attributes simultaneously. Beyond estimating main effects, scholars increasingly analyze heterogeneity through subgroup analysis and contextual variables, raising methodological challenges in detecting and interpreting interaction effects. Statistical power constraints, common in survey experiments, further complicate this task. This paper addresses the question: how can both main and interaction effects be reliably inferred in conjoint studies? We contribute in two ways. First, we conduct a systematic evaluation of leading approaches—including post-hoc corrections, sparse regression methods, and Bayesian models—across simulation regimes that vary sparsity, noise, and data availability. Second, we propose a novel black-box inference framework that leverages machine learning to recover main and interaction effects in conjoint experiments. Our approach balances computational efficiency with accuracy, providing a practical tool for researchers studying heterogeneous effects.
\end{abstract}

\section{Introduction}
Since their formal introduction to the community in the 2010s \citep{hainmueller2014causal}, conjoint analysis survey designs have continued to permeate survey research in political science \citep{zhirkov2022estimating} and elsewhere in the social sciences \citep{agarwal2015interdisciplinary}. At a high level, a \emph{conjoint experiment} is a survey instrument that can be used to assess survey respondent preferences across a number of attributes simultaneously. This differs from a traditional survey experiment, which tends to manipulate one (or at most a few) variables at once. The ability to examine how multiple factors simultaneously influence respondents’ decisions, on topics such as immigration \citep{hainmueller2014causal}, who to vote for \citep{green2022strategic}, and policy preferences \citep{horiuchi2018measuring}, together with favorable properties relevant to the impact of social desirability bias \citep{horiuchi2022does}, make conjoint designs an important area of methodological development.

In addition to studying the causal impact of a particular set of variables on survey respondent decisions, scholars also commonly use results from conjoint experiments to explore patterns in how these effects vary across subpopulations, or what is called \emph{subgroup analysis}. For instance, \citet{bansak2016economic} group participants based on their political alignment, age, level of education, and income, and subsequently study their attitude towards asylum seekers in Europe. In addition to subgroup analysis, scholars have also begun to explore how auxiliary experimental variables, or what we will call (decision) \emph{context variables}, impact responses. Context variables are factors that describe the circumstances or environment in which the experiment takes place, and may influence participant responses. As one recent example, \citet{green2022strategic} experimentally manipulate the question framing they use in a conjoint experiment on voting behavior.

Context variables and subgroup analyses are both cases where researchers are interested in interaction effects between variables \emph{within} the conjoint experiment and variables \emph{external} to the conjoint portion of the study. As \citet{leeper2020measuring} point out, there are at least two reasons to be careful in the analysis of such interaction effects. First, without care to analyze the appropriate marginal quantities, one can easily misinterpret the absolute impacts of a particular conjoint variable across sub-populations or context levels. Second, without explicit consideration of a statistical model with interaction terms, researchers can end up trying to provide interpretations of differences across subgroups that are not statistically meaningful. To address the latter concern, \citet{leeper2020measuring} propose to formally test for interaction effects before interpretation by comparing two models: one with main effects of conjoint variables, and one with interactions between conjoint and subgroup variables. However, conjoint experiments are ultimately survey experiments and are thus subject to concerns regarding statistical power. These concerns are only exacerbated when analyzing and comparing subgroups, motivating the primary research question pursued in this study: \emph{How can both main effects and subgroup/context interaction effects be reliably detected in conjoint experiments?} 

A number of solutions have been proposed to tackle the problem of identifying interaction terms in potentially underpowered settings \citep{bien2013lasso,lim2015learning,blackwell2022reducing}, including in  conjoint experiments \citep{ratkovic2017sparse,liu2023multiple}.
Each of these existing approaches come with strengths and weaknesses that vary according to properties of the data. \textit{One goal of the present work} is therefore to compare the effectiveness of these methods in identifying interaction effects in conjoint experiments under (i) varying levels of sparsity in main and interaction effects, (ii) noise assumptions, and (iii) regimes of data availability. To do so, we  extend and synthesize recently proposed methods for simulating conjoint experimental data \citep{stefanelli2020subjects,liu2023multiple,schuessler2020power,gall2020simulation} and provide an extensive performance comparison of existing methods. \emph{The second goal of this work} is to develop, and to evaluate in the context of these existing methods, a new approach for inferring main and subgroup (or context) interaction effects for conjoint analysis. Our approach makes use of both the novel setup of conjoint analysis, and recent advancements in the use of black-box models to perform statistical inference \citep{jerzak2023improved, teblunthuis2024misclassification, angelopoulos2023prediction, chernozhukov2018double} to construct a simple and effective method for inferring subgroup and contextual interaction effects. 
Compared to existing methods, our approach offers a practical compromise between computational efficiency and accurate recovery of interaction effects under moderate sparsity. In summary, our work makes the following contributions to the literature:
\begin{itemize}
    \item We present a comprehensive empirical evaluation of leading approaches, including frequentist, Bayesian, and machine learning-based methods, for identifying heterogeneous effects in conjoint experiments. Our analysis highlights key trade-offs and performance comparisons across methods.
    \item We propose a novel encoding technique and black-box framework  that achieves low false positive rates while reliably detecting true interaction effects under reasonable assumptions on the sparsity of factors that affect decision-making in binary forced choice conjoint experiments. 
    \item We develop new, extendible simulation code designed for benchmarking methods that estimate main and interaction effects in conjoint designs. This enables both replication and future methodological extensions.
\end{itemize}


\section{Literature Review}

A number of recent works have focused on potential methodological pitfalls in conjoint experiments, from issues with common statistical estimators (i.e. problems in using the Average Marginal Component Effect, or AMCE) \citep{zhirkov2022estimating,myers2024welfare}, to problems in assuming independent effects of conjoint factors \citep{egami2019causal}, to problematizing assumptions about the lack of ordering effects when showing multiple tasks to participants \citep{rudolph2024ordering}. While these issues are critical, they are separate from the question pursued here, which focuses on the measurement of interaction effects between conjoint profile variables and either respondent subgroups or decision context variables. As such, we set aside discussion of these other important issues with conjoint methods unless directly relevant to the question at hand.

\subsection{Inferring Interaction Effects (in Conjoint Experiments)}

The core focus of our work is on measuring the existence and magnitude of interaction effects between conjoint and subgroup/context variables in conjoint experiments.  We identify four representative approaches for doing so in the literature: one that is widely used in practice \citep{leeper2020measuring} and a follow-up that uses post-hoc corrections to this standard model \citep{liu2023multiple}, as well as more complex Bayesian \citep{ratkovic2017sparse} and optimization-based \citep{bien2013lasso} models that are directly applicable to the problem at hand.

Perhaps the most well known work to focus on measuring interaction effects in conjoint experiments is from \citet{leeper2020measuring}, who point out the common pitfalls of using differences in AMCEs to understand differences in preferences across subgroups and present a simple but effective solution. Specifically, the authors propose to use an omnibus $F$-test, which leverages two models, one with interaction terms between conjoint features and subgroup identifiers and one without, to determine the significance of the conjoint features. While straightforward, \citet{liu2023multiple} note that such an approach does not work well in high-dimensional settings, particularly with smaller sample sizes associated with survey designs, because assumptions about sample size break down, models become over-parameterized, and false positives are likelier to occur at a higher rate.  To address this, \citet{liu2023multiple} assess the effectiveness of standard $p$-value correction techniques using simulated conjoint data. In particular, the authors assess three $p$-value correction methods: Bonferroni, Benjamini-Hochberg, and adaptive shrinkage, finding that adaptive shrinkage should be used as the default choice when there is limited information about the existence of AMCEs.

Others have opted to make use of modeling, rather than post-hoc corrections, to address challenges with estimating interaction effects between subgroups and conjoint variables. \citet{ratkovic2017sparse} construct a Bayesian model named \texttt{LassoPlus}, which uses lasso-inspired priors and an informed data post-processing step to address the challenge of selecting interaction effects alongside the main effects of conjoint variables. While such a model can theoretically address the core task of interest in this paper, in practice, Bayesian models are subject to known challenges in parameter estimation that may limit their effectiveness. Moreover, it is not immediately obvious whether this approach is able to identify effects in settings where the sparsity assumption is relaxed (i.e., where there are a significant number of interaction effects). 

External to the study of conjoint models, scholars have developed purely optimization-focused adaptations of the lasso that address the same challenge of estimating interaction effects in regression models in small-$N$ datasets. Most notably, \citet{bien2013lasso} develop \texttt{LassoNet} which formulates interaction models using lasso-style constraints on a linear model in a weak hierarchical setting, where an interaction term exists when at least one of the main effects is present in the model. By setting constraints on the magnitude of the main effects that constitute the interaction terms, the interaction matrix satisfies the hierarchical constraint, and can be solved using standard optimization procedures similar to those used for optimization in the standard lasso.

The approaches of \citet{leeper2020measuring}, \citet{liu2023multiple}, \citet{ratkovic2017sparse}, and \citet{bien2013lasso} thus serve as an incomplete but largely representative sample of possible approaches to addressing questions about how to measure the existence and magnitude of interaction effects between conjoint and subgroup/context variables in conjoint experiments. We select these four methods because they span a range of modeling strategies, frequentist, Bayesian and machine learning, and are widely cited in the literature. We adopt these methods as competitors to our new approach, and assess the effectiveness of each method along a variety of dimensions and in a variety of assumed properties of the data-generating distribution.

\subsection{Using Black-box models to Perform Statistical Inference}

Our proposed approach relies on the principled use of a black-box machine learning model to perform statistical inference, and is motivated by several recent lines of research in this domain. First, our work engages in the discussion on the repercussions of, and solutions to, the problem of treating machine learning models as unbiased classifiers and making statistical inferences in downstream tasks. 
Most relevant is work focusing on leveraging both labeled and unlabeled data to obtain tighter confidence intervals compared to using only labeled data \citep{angelopoulos2023prediction,zrnic2024cross,rister2025correcting}. Though our method serves a similar purpose, the binary forced choice conjoint design presents its own unique challenges, such as the prevalence of measurement error and the limited information researchers can obtain from them (compared to numerical ratings).  

Also relevant is work in causal estimation is debiased or double machine learning (DML), as introduced by \citet{chernozhukov2018doub}. 
DML applies cross-fitting to nuisance estimation, which requires applying it separately to a large number of interaction terms. This can become computationally burdensome. 
Importantly, however, conjoint experiments involve \emph{randomized assignment} of attributes and contexts, which eliminates confounding by design and provides known assignment probabilities. This removes the need to estimate inverse propensity scores, allowing for (with our setup) more efficient design-based inference.  Although ML-based estimators may still introduce regularization bias, we posit that by explicitly incorporating the known randomization structure, we can mitigate such biases without the full complexity of the DML framework.

Specifically in the context of conjoint analysis, two studies in particular are worth noting. \citet{grimmer2017estimating} argue that it could be challenging for researchers to select a method to estimate treatment effects that suit their dataset. 
However, this approach does not take into account the structure of conjoint analysis with context, which omits certain interaction terms. By making use of this structure, we are able to provide a narrower but more effective solution to the focal problem of this paper. \citet{ham2024using} are also interested in a slightly different question than the one posed here. They aim to identify whether a particular factor is significant given the other factors. Their assumption-free approach can identify significant effects as well as test the validity of commonly invoked assumptions in conjoint analysis (e.g., no profile order effect). Nonetheless, their approach is more suitable for early-stage exploratory analysis, as it does not measure the direction or magnitude of the effects.

\subsection{Simulating Conjoint Experiment Data}
 Two recent studies have run simulations to conduct power analysis in conjoint settings. \citet{stefanelli2020subjects} use simulation to offer insights into the limitations of experimental design in conjoint analysis, and \citet{schuessler2020power} aim to robustly estimate type 1 and type 2 errors as well as calculate the minimal number of respondents needed to obtain significant AMCEs. We draw on and extend their work here.

\section{Notation and Problem Statement}

\subsection{Overview}
We assume a binary forced choice design, perhaps the most common setup in conjoint experiments. In a forced choice design, survey participants are presented with two options, each a random selection of conjoint experiment variables, and asked to choose one of the two profiles presented to them. Hence, the output variable, $Y$, is binary. There are two sets of independent variables in our model: $X$ and $C$, where $X$ denotes the set of conjoint attributes, and $C$ denotes the set of context attributes or subgroup identifiers. Notably, then, we lump these two together for our analysis for simplicity.  The terms context and subgroup are used interchangeably throughout the rest of this study. $B_j$ denotes the coefficient of conjoint attribute $X_{j}$, where $j \in \{1, \dots, N\}$ indexes the $N$ unique one-hot encoded conjoint attributes and $i \in \{1, \dots, N_s\}$ indexes observations. $B_{j,k}$, or $B_z$ for simplicity, denotes the coefficient of the interaction term between the conjoint variable $X_{j}$ and the context variable $C_{k}$, where $k \in \{1, \dots, M\}$ indexes the $M$ unique one-hot encoded context variables (e.g., subgroup identifiers). The response variable $Y_i$ corresponds to the outcome for observation $i$. The goal in our setting is to estimate $B_{j} \in \mathbb{R}^N$ and $B_{j,k} \in \mathbb{R}^{N \times M}$. To avoid introducing new variables, $X_j$ denotes the variable, and $X_{i,j}$ represents $X_j$ at $i$--th observation; the same principle applies to $C$.

\begin{equation} 
Y_i = \sum_{j=1}^{n} B_{j}X_{i,j} + \sum_{k=1}^{m}\sum_{j=1}^{n} B_{j,k}X_{i,j}C_{i,k} + \epsilon
\label{eq:model}
\end{equation}

Finally, while we discuss it in more detail below, we note here that our simulation model, given this setup, is summarized in Equation~\eqref{eq:model}.

\subsection{Quantity of Interest}
Numerous causal interaction measures have been proposed in the literature \citep{vanderweele2015explanation,dasgupta2015causal,egami2019causal}, most of which focus on interactions between \emph{attributes} in standard conjoint designs. At first glance, one might consider extending existing estimands such as the Average Interaction Effect (AIE) \citep{vanderweele2015explanation,dasgupta2015causal} or the Average Marginal Interaction Effect (AIME) \citep{egami2019causal} to include context variables by treating them as additional attributes. However, this approach breaks key design assumptions of our setup.

Specifically, by construction, context variables are held constant across each profile pair. Incorporating them as if they were independently randomized attributes violates the foundational principle of random assignment in conjoint analysis. Additionally, many existing methods assume the existence of the average marginal effect (AME) \citep{hainmueller2014causal} for each variable. Since context variables do not vary within a profile pair, this assumption does not hold, rendering interaction estimates involving context invalid or uninterpretable.

This motivates the need for a distinct estimand tailored to attribute--context interactions. We borrow some notation from \citet{egami2019causal} to define our quantity precisely. Formally, consider a conjoint design with \( N' \) factorial conjoint treatments  and \( M' \) context treatments. Let the factorial and context treatments have \( L_n \) and \( L_m \) levels, respectively. After one-hot encoding, there are \( N = N' \times L_n \) factorial conjoint variables, denoted by \( X_i = (X_{i,1}, X_{i,2}, \dots, X_{i,N}) \), and \( M = M' \times L_m \) factorial context variables, denoted by \( C_i = (C_{i,1}, C_{i,2}, \dots, C_{i,M}) \). Let \( X_{i,{1:K}} \) denote a subset of \( K \) treatment variables and \( X_{i,{K+1:N} }\) the remainder; the same notation applies to \( C \). 

We define our quantity as follows:

\begin{align}
B_{1,1} \equiv \mathbb{E} \Bigg[ \, & 
 \int \big( Y_i(X_{i,{1}} = 1, X_{i,{2:N}}) \notag \\ 
& -  Y_i(X_{i,{1}}  = 0, X_{i,{2:N}})\big) \, dF(X_{i,{2:N}}) \notag \\
& - \int \big( Y_i(X_{i,{1}} = 1, X_{i,{2:N}}, C_{i,{1}} = 0) \notag \\
& +  Y_i(X_{i,{1}} = 0, X_{i,{2:N}}, C_{i,{1}} =0) \,\big) \, dF(X_{i,{2:N}}, C_{i,{2:M}})
\Bigg]
\label{eq:rho_definition}
\end{align}

\noindent This quantity captures the interaction term between attribute level \( X_1 \) and context \( C_1 \). If the baseline condition \( (X_k, C_0) \) has zero effect, then $B_{1,1}$ can be interpreted causally. If not, one can difference out the baseline by computing $B_{1,1} - B_{k,0}$. Since the choice of \( X_k \) is arbitrary, this quantity is invariant to the selection of attribute baselines. 

Once interaction terms $B_{1,m}$for all \( m \in \{1, \dots, M\} \) are available, we can identify the AME of \( X_1 \) by algebraic decomposition:

\begin{align}
B_{1} \equiv\; & \mathbb{E} \Bigg[ 
    \int Y_i(X_{i,{1}} = 1, X_{i,{2:N}})  \notag \\
& \quad -Y_i(X_{i,{1}} =0, X_{i,{2:N}}) \, dF(X_{i,{2:N}}) 
\Bigg] 
 - \sum_{m=1}^{M} B_{1,m}
\label{eq:tau_definition}
\end{align}

\section{A New Method for Identifying Attribute-Context Interaction Effects}

\subsection{An Initial Model}
\label{sec:derivation}

Our new approach estimates main effects and interaction terms by manipulating the expected values of selected variables and computing their differences. To obtain these expectations, we train a machine learning model that captures the nonlinear relationships between conjoint attributes, context variables, and the response. This allows us to marginalize over certain variables by computing conditional expectations. For example, $\mathbb{E}[Y \mid X_k = 1]$  denotes the expected outcome when $X_k =1$, obtained by using the trained black-box model to predict outcomes on the test data where $X_k =1$, thereby averaging over the distribution of all other variables. 

We employ a multilayer perceptron (MLP) as the black-box model for estimating these expectations. The model takes one-hot encoded conjoint attributes and context variables as input features, and the response is a binary indicator of whether a profile is selected. After training, we perform a series of prediction tasks, described below, to recover the estimated coefficients corresponding to each variable. 

In order to calculate the coefficient of an arbitrary main effect $X_k$ and all the interaction terms associated with it, we need to calculate a few expected values. The first set of expected values, $\mathbb{E}[Y \mid X_k = 1]$~\eqref{eq:xk1} and $\mathbb{E}[Y \mid X_k = 0]$~\eqref{eq:xk0}, is processed by partitioning the test set into two parts: $X_k = 1$ and $X_k = 0$. Upon partitioning the data, we use the trained model to make predictions on the two test sets and compute their means. The mean is then interpreted as the expected value. Subtracting the mean of the $X_k = 0$ subset from the $X_k = 1$ subset yields $\mathbb{E}[Y \mid X_k = 1] - \mathbb{E}[Y \mid X_k = 0]$.

\[
\text{Let } k \in \{1, \ldots, n\}
\]

\begin{equation}
\mathbb{E}[Y \mid X_k = 1] = B_k + \sum_{i=1}^{n \setminus k} B_i \mathbb{E}[X_i] + \sum_{j=1}^{m} B_{k,j} C_j + \sum_{j=1}^{m} \sum_{i=1}^{n \setminus k} B_{i,j} \mathbb{E}[X_i] C_j
\label{eq:xk1}
\end{equation}

\begin{equation}
\mathbb{E}[Y \mid X_k = 0] = \sum_{i=1}^{n \setminus k} B_i \mathbb{E}[X_i] + \sum_{j=1}^{m} \sum_{i=1}^{n \setminus k} B_{i,j} \mathbb{E}[X_i] C_j
\label{eq:xk0}
\end{equation}

\begin{equation}
\mathbb{E}[Y \mid X_k = 1] - \mathbb{E}[Y \mid X_k = 0] = B_k + \sum_{j=1}^{m} B_{k,j} C_j
\label{eq:xk1-xk0}
\end{equation}

Equation~\eqref{eq:xk1-xk0} captures the coefficient for main effect $k$ and all associated interaction terms with $X_k$. Let $l \in \{1, \ldots, m\}$. To isolate the coefficient $B_{k,l}$, we compute $\mathbb{E}[Y \mid X_k = 1, C_l = 0] - \mathbb{E}[Y \mid X_k = 0, C_l = 0]$.

\begin{equation}
\mathbb{E}[Y \mid X_k = 1, C_l = 0] = B_k + \sum_{i=1}^{n \setminus k} B_i \mathbb{E}[X_i] + \sum_{j=1}^{m \setminus l} B_{k,j} C_j + \sum_{j=1}^{m \setminus l} \sum_{i=1}^{n \setminus k} B_{i,j} \mathbb{E}[X_i] C_j
\label{eq:xk1cl0}
\end{equation}

\begin{equation}
\mathbb{E}[Y \mid X_k = 0, C_l = 0] = \sum_{i=1}^{n \setminus k} B_i \mathbb{E}[X_i] + \sum_{j=1}^{m \setminus l} \sum_{i=1}^{n \setminus k} B_{i,j} \mathbb{E}[X_i] C_j
\label{eq:xk0cl0}
\end{equation}

\begin{equation}
\mathbb{E}[Y \mid X_k = 1, C_l = 0] - \mathbb{E}[Y \mid X_k = 0, C_l = 0] = B_k + \sum_{j=1}^{m \setminus l} B_{k,j} C_j
\label{eq:xk1cl0-xk0cl0}
\end{equation}

Equation~\eqref{eq:xk1cl0-xk0cl0} is similar to Equation~\eqref{eq:xk1-xk0} but excludes the interaction term $B_{k,l} C_l$. Subtracting Equation~\eqref{eq:xk1cl0-xk0cl0} from Equation~\eqref{eq:xk1-xk0} isolates the interaction:

\begin{equation}
\eqref{eq:xk1-xk0} - \eqref{eq:xk1cl0-xk0cl0} = B_k + \sum_{j=1}^{m} B_{k,j} C_j - \left(B_k + \sum_{j=1}^{m \setminus l} B_{k,j} C_j \right) = B_{k,l} C_l
\label{eq:two_exp_sub}
\end{equation}

Since $C_l$ is fixed, we compute it based on the known distribution of context. Dividing both sides of Equation~\eqref{eq:two_exp_sub} by $C_l$ yields the estimated coefficient $B_{k,l}$.

\begin{equation}
\sum_{j=1}^{m} \left( \mathbb{E}[Y \mid X_k = 1] - \mathbb{E}[Y \mid X_k = 0] - \left( \mathbb{E}[Y \mid X_k = 1, C_j = 0] - \mathbb{E}[Y \mid X_k = 0, C_j = 0] \right) \right) = \sum_{j=1}^{m} B_{k,j} C_j
\label{eq:sum_int}
\end{equation}

Equation~\eqref{eq:sum_int} represents the total contribution of interaction terms for variable $X_k$, which corresponds to the second term in Equation~\eqref{eq:xk1-xk0}. Therefore, subtracting Equation~\eqref{eq:sum_int} from Equation~\eqref{eq:xk1-xk0} yields the coefficient $B_k$:

\begin{equation}
\eqref{eq:xk1-xk0} - \eqref{eq:sum_int} = \left( B_k + \sum_{j=1}^{m} B_{k,j} C_j \right) - \sum_{j=1}^{m} B_{k,j} C_j = B_k
\label{eq:last}
\end{equation}

\subsection{Extending the Approach with Difference Encoding}
In a forced-choice conjoint design, respondents evaluate a pair of profiles (for simplicity, here, a ``left'' and ``right'' profile) and select the one they prefer. Each profile is treated as an independent observation, and a label of 1 indicates that the profile was selected in a given task. This approach ignores the dependency between profiles within the same pair, and thus does not explicitly account for the pairwise structure inherent in the design. However, it is reasonable to expect that explicitly modeling the difference between paired profiles might enhance predictive performance and, consequently, improve inference \citep{grimmer2017estimating}. To this end, we consider an extension of our model where rather than treating each profile separately, we represent each pair as a single observation by subtracting the left profile's attributes from the right's. Specifically, we perform one-hot encoding for all attribute levels and compute a difference vector where each element takes on one of three values: 1 (the attribute is present in the left profile  only), $-1$ (present in the right profile only), or 0 (present or absent in both profiles). Formally, this can be expressed as:
\[
\mathbf{X}_i^{(D)} = \left( X_{i,1}^L - X_{i,1}^R,\, \ldots,\, X_{i,N}^L - X_{i,N}^R \right).
\]

The context variables are kept unchanged, as both profiles in a given pair are presented under the same scenario. This setup, we posit, allows the model to better learn interactions between attribute differences and contextual features. The binary response variable is encoded as 1 if the left profile is chosen, and 0 if the right is chosen.

We refer to this modeling strategy as \emph{difference-encoded predictive causal estimation (\texttt{DiPCE})}. While this subtraction-based representation is commonly used in simulation pipelines to generate labels, \texttt{DiPCE} applies it during model training to exploit the pairwise structure for both prediction and inference. One caveat is that one-hot encoding eliminates the exclusivity of levels within a categorical attribute (e.g., only one education level is valid per profile); this constraint must be implicitly learned by the model. Details on how we approach estimation for the \texttt{DiPCE} model is provided in Appendix C in the supplementary materials.

\section{Evaluation Approach}

\subsection{Simulation Model}
Our simulation is based on an influential study in causal inference for conjoint analysis, which investigates the attributes of immigrants that influence support for their admission \citep{hainmueller2014causal}.
The original design includes the following nine attributes: gender, education, language, country of origin, profession, job experience, job plans, reason for application, and prior trips to the U.S. These attributes have 7, 2, 10, 4, 3, 11, 4, 4, and 5 levels, respectively, totaling 50 levels across all attributes. Each immigrant profile is constructed by independently and randomly drawing one level for each attribute. 

To extend the original design, we incorporate five additional context variables that serve as alternative framings of the decision-making scenario. With 50 attribute levels, this results in 50 potential main effects. Each level may also interact with the five context variables, leading to 250 possible interaction terms, for a total of 300 potential effects. Recognizing that it is unrealistic for all 300 effects to be non-zero in empirical studies, we introduce varying sparsity levels. Specifically, we apply four levels of sparsity, 0.5, 0.65, 0.8, and 0.95, to both main and interaction effects. For example, a sparsity setting of 0.5 means that 50\% of the effects are randomly set to zero. However, we assume that main effects should not be sparser than interaction effects, and we therefore restrict our design to configurations where the interaction sparsity is greater than or equal to the main sparsity. 

In each simulation, 2,000 respondents evaluate 8 paired profiles (i.e., 16 total profiles per respondent). With 10 sparsity configurations and 15 replications per configuration, this yields a total of 2 profiles × 8 tasks × 2,000 respondents × 10 configurations × 15 replications = 4,800,000 data points split across the fifteen replications. Replications are used to estimate standard errors on performance metrics. 

After generating all variables, we fix the first level of each conjoint attribute to zero in order to ensure identifiability of the model. For illustration, consider the attribute \textit{Job Experience}, which has four levels: ``none,'' ``1--2 years,'' ``3--5 years,'' and ``5+ years.'' Although there are four levels (and hence four potential coefficients), we fix the effect of the baseline level (``none'') to zero. This approach is applied across all nine conjoint attributes, resulting in 9 out of the 50 main effect coefficients being set to zero. Similarly, for the 250 interaction terms (each of the 50 attribute levels crossed with 5 context variables), we fix the interactions involving these baseline levels to zero, leaving 164 (41 attributes * 4 contexts) potentially non-zero interaction terms. Next, we apply the sparsity settings described earlier. For the remaining (non-fixed) main and interaction effects, we assign coefficient values by drawing from a uniform distribution, 
$\text{Unif}(-1, 1) \setminus (-0.1, 0.1),$
which excludes the interval $(-0.1, 0.1)$ to avoid generating negligible effects that are difficult to detect. A detailed description of the parameters is provided in Table~\ref{table:parameter}.

\begin{table}
\centering
\footnotesize
\caption{\label{table:parameter}Tabular description of parameters involved in the construction of our simulation model. }

\begin{tabular} {||p{3cm}| p{3cm} | p{3cm}| p{3cm} ||}
\hline
Parameter & Values Taken  & Parameter & Values Taken \\
\hline \hline
 $\mathcal{C}$: Levels of context list & $[5]$ &
$\mathcal{M}$: Levels of conjoint list & $[7, 2, 10, 4, 3, 11, 4, 4, 5]$\\ 
\hline
$Sp_{j}$: Sparsity options for main effects  & $[0.5, 0.65, 0.8, 0.95]$  & 
$Sp_{z}$: Sparsity options for interaction effects  & $[0.5, 0.65, 0.8, 0.95]$ \\

\hline
$M_{e}$: Measurement error  & $85\%$ & 
$Q$: Number of questions per respondent  & $8$ \\ 
\hline

$B_{j}$: Coefficients of main effects  & $\text{Unif}(-1, 1) \setminus (-0.1, 0.1)$ & 
$B_{z}$: Coefficients of interaction  & $\text{Unif}(-1, 1) \setminus (-0.1, 0.1)$ \\ 
\hline

$N_{r}$: Number of simulations  & $15$ & 
$N_{s}$: Number of respondents per simulation  & $2000$ \\ 
\hline

\hline
\end{tabular}
\end{table}

In the conjoint setting, the outcome variable reflects the profile selected by the respondent from a pair. In our simulation, respondents ``choose'' a profile by comparing latent utility scores, which are computed using the profile’s attribute values and the assigned coefficients. Specifically, the score for each profile is calculated as:
\[
\text{score} = \mathbf{x}^\top \boldsymbol{\beta} + (\mathbf{x}^\top \mathbf{B}) \cdot \mathbf{z},
\]
where $\mathbf{x}$ is the vector of profile attributes, $\boldsymbol{\beta}$ is the vector of main effects, $\mathbf{B}$ is the interaction coefficient matrix, and $\mathbf{z}$ is the context vector. To make the simulation more realistic, we introduce stochasticity via measurement error. This error reflects the probability that a respondent does not choose the profile with the higher computed score. Based on findings from \citet{clayton2023correcting}, we assume a measurement error rate of approximately 15\%. For each choice, we draw a value $ME \sim \text{Unif}(0, 1).$
If $ME > 0.85$, the respondent's decision is flipped, that is, they select the lower-scoring profile instead. Our primary quantities of interest include the main effects (e.g., Gender, Education Level) and their interactions with context variables (e.g., Gender $\times$ Context1, Education Level $\times$ Context2). For each effect, we evaluate whether it is null, and if not, whether the estimated direction (positive or negative) matches the true sign. A detailed description of the simulation generation process is provided in Algorithm ~\ref{algo:simulation}.

\begin{algorithm}[t] 
\begin{multicols}{2}
\footnotesize
\SetAlgoLined
\DontPrintSemicolon
\SetInd{0.5em}{1.2em}

\KwIn{ $\mathcal{C}$, $\mathcal{M}$, $Sp_{j}$, $Sp_{z}$, $N_{s}$, $N_{r}$, $Q$, $M_{e}$}
\KwOut{$\text{Results} \gets \text{list of } (q_0, q_1, \text{RespondentId}, a^\prime)$}

\textbf{Initialize } $\text{Results} \gets [\,]$ \;

\For{$sp_{j} \in \mathcal{S}p_{j}$}{
  \For{$sp_{z} \in \mathcal{S}p_{z}$}{
    \For{$n_{s} \in \mathcal{N}_{s}$}{
      $B_{j}^{sp_{j}} \gets \text{FixedFirstLevel}(\text{random}(\sum \mathcal{M}))$ \;
      $B_{z}^{sp_{z}} \gets \text{FixedFirstLevel}(\text{random}(\sum(\mathcal{M}, \mathcal{C})))$ \;
      
      \For{$n_{r} \in \mathcal{N}_{r}$}{
        $c_{t} \gets \text{random}(\mathcal{C})$ \;
        
        \For{$q \in \mathcal{Q}$}{
          $q_0 \gets \text{random}(\text{Question}(\mathcal{M}))$ \;
          $q_1 \gets \text{random}(\text{Question}(\mathcal{M}))$ \;
          $y_0 \gets q_0 \cdot B_{j}^{sp_{j}} + (q_0 \cdot B_{z}^{sp_{z}}) \cdot c_t$ \;
          $y_1 \gets q_1 \cdot B_{j}^{sp_{j}} + (q_1 \cdot B_{z}^{sp_{z}}) \cdot c_t$ \;
          $a \gets \arg\max(y_0, y_1)$ \;
          $p \gets \text{uniform}(0,1)$ \;
          
          \uIf{$p \ge M_e$}{
            $m \gets 1$ \;
          }\Else{
            $m \gets 0$ \;
          }
          
          $a^\prime \gets \text{XOR}(a, m)$ \;
          \textbf{Append } $(q_0, q_1, \text{RespondentId}, a^\prime)$ \textbf{ to } $\text{Results}$ \;
        }
      }
    }
  }
}

\caption{One Run of Simulation}
\label{algo:simulation}
\columnbreak

\textbf{Notations:} \;
\vspace{0.5em}

$\mathcal{C}$: Context List \\
$\mathcal{M}$: Conjoint List \\
$Sp_{j}$: Sparsity of main effects \\
$Sp_{z}$: Sparsity of interaction effects \\
$N_{s}$: Number of simulations \\
$N_{r}$: Number of respondents \\
$Q$: Number of questions per respondent \\
$M_{e}$: Measurement error probability \\
$\text{Results}$: Output data list containing tuples of $(q_0, q_1, \text{RespondentId}, a^\prime)$

\end{multicols}
\end{algorithm}

\subsection{Outcome Metrics}

\subsubsection{Coefficient Recovery}
We first evaluate model performance from the perspective of a prediction problem on the ability of the model to identify real positive, real negative, or zero effects for each of the main and interaction effects. That is, because our task involves directional classification, each coefficient is assigned one of three labels: positive (``$+$''), negative (``$-$''), or null (``$0$''). Accuracy is then defined as the proportion of coefficients that are identified as being in in one of these three categories, where that category matches the truth from the simulation model. This metric captures how effectively the model can identify both real and non-zero effects, overall.

The definitions of \textbf{false positive rate (FPR)}, and \textbf{false negative rate (FNR)} for binary settings are:
\[
\text{FPR} = \frac{\text{False Positives (FP)}}{\text{False Positives (FP)} + \text{True Negatives (TN)}}, \quad
\text{FNR} = \frac{\text{False Negatives (FN)}}{\text{False Negatives (FN)} + \text{True Positives (TP)}}.
\]

We then redefine TP, FP, FN, and TN for this multiclass setting as follows: 

\begin{itemize}
  \item \textbf{True Positives (TP)}: Number of non-zero coefficients correctly predicted with the correct sign.
  \[
  \text{TP} = \sum (\texttt{pred} = \texttt{truth} \neq 0)
  \]

  \item \textbf{False Positives (FP)}: Includes (a) coefficients predicted as non-zero when the true value is zero, and (b) coefficients predicted with the incorrect non-zero direction.
  \[
  \text{FP} = \sum (\texttt{pred} \neq 0,\, \texttt{truth} = 0) + \sum (\texttt{pred} \neq 0,\, \texttt{truth} \neq 0,\, \texttt{pred} \neq \texttt{truth})
  \]

  \item \textbf{False Negatives (FN)}: Number of true non-zero coefficients incorrectly predicted as zero.
  \[
  \text{FN} = \sum (\texttt{pred} = 0,\, \texttt{truth} \neq 0)
  \]

  \item \textbf{True Negatives (TN)}: Number of coefficients correctly predicted as null.
  \[
  \text{TN} = \sum (\texttt{pred} = \texttt{truth} = 0)
  \]
\end{itemize}

This multiclass framework allows us to evaluate not only whether an effect is detected, but also whether its direction is accurately inferred.

\subsubsection{Predictive Accuracy}

We also consider how well the model can predict a held-out sample of (simulated) survey respondent answers. While the primary goal of this study is coefficient estimation rather than prediction, predictive performance in a traditional machine learning setting still offers valuable insights. Accuracy serves as an indicator of how well a model learns the underlying data-generating process. By comparing accuracy on the training and test sets, we can assess whether the model is capturing true signal or overfitting to noise, akin to evaluating the bias-variance tradeoff.

Using predictive performance to inform inference is not without precedent. For example, \citet{grimmer2017estimating} employ an ensemble method to estimate heterogeneous treatment effects, combining estimates from multiple models including black-box approaches such as random forests. Each model contributes a coefficient estimate, and the final estimate is a weighted average where the weights are based on each model’s out-of-sample predictive performance. Their rationale is that models with stronger predictive performance are better able to uncover systematic heterogeneity in treatment effects, leading to more accurate inference.

We train an MLP model (refer to Appendix B in the supplementary materials for full architecture details) alongside baseline models such as linear regression, \texttt{HierNet}, and \texttt{LassoPlus} on the training set and evaluate their performance on the test set.  For models with built-in binary classification functionality, such as \texttt{HierNet} and MLP, we directly apply the inbuilt \texttt{predict} function. For models lacking native binary classification output, specifically linear regression and \texttt{LassoPlus}, we compute predicted values using the estimated coefficients; predictions exceeding 0.5 are classified as 1, and 0 otherwise. We also report the accuracy of the MLP both with and without difference encoding to assess the value of this feature representation. The full set of predictive accuracy results is presented in Section~\ref{sec:predicted_accuracy_results}.

\subsection{Baseline Models}
Our baseline models are drawn, as noted above, from prior work. We  linear regression (\texttt{lm}) as a standard approach, two p-value correction approaches from \citet{liu2023multiple} (linear regression with adaptive shrinkage correction, \texttt{lm-ash}, and with Bonferroni correction, \texttt{lm-bon}), \texttt{HierNet} \cite{bien2013lasso}, and \texttt{LassoPlus} \cite{ratkovic2017sparse}. 

The posterior distribution in Ash is given in \eqref{eq:posterior}, where $\beta$ denotes the effects of interest, and $\hat{\beta}$ and $\hat{s}$ represent the estimated effect size and standard error, respectively.
\begin{equation}
p(\beta \mid \hat{\beta}, \hat{s}) \propto p(\hat{\beta} \mid \beta, \hat{s}) \, p(\beta \mid \hat{s}).
\label{eq:posterior}
\end{equation}
The core idea of \eqref{eq:posterior} is to compute the likelihood of $\hat{\beta}$ being from the same distribution as  $\beta$ with standard error $\hat{s}$. 

\texttt{HierNet} assumes a two-way interaction model, \eqref{eq:hierN}, for a dependent variable Y and independent variable X.
\begin{equation}
Y = \beta_0 + \sum_j \beta_j X_j + \frac{1}{2} \sum_{j \ne k} \Theta_{jk} X_j X_k + \varepsilon,
\label{eq:hierN}
\end{equation}
$\beta$ denotes the coefficient of $X$, and $\Theta$ denotes the coefficient of interaction terms $X_j*X_k$, for $j  \neq k$. The simplified version of \texttt{HierNet}’s overarching goal is to minimize the loss function, the sum of the loss function $q$, and the regularization term $\lambda$, as shown in 

\begin{equation}
\begin{array}{ll}
\displaystyle \min_{\substack{\beta_0 \in \mathbb{R},\ \beta \in \mathbb{R}^p,\\ \Theta \in \mathbb{R}^{p \times p},\ \Theta = \Theta^T}} 
& q(\beta_0, \beta, \Theta) + \lambda \sum_j \max\{ |\beta_j|, \lVert \Theta_j \rVert_1 \} + \frac{\lambda}{2} \lVert \Theta \rVert_1
\end{array}
\label{eq:minimization}
\end{equation}

In order to satisfy weak hierarchy, they remove the symmetry constraint on the interaction coefficient, $\Theta$, such that
\[
\lVert \Theta_j \rVert_1 \leq \beta_j^{+} + \beta_j^{-} 
\quad \text{and} \quad 
\beta_j^{+} \geq 0,\ \beta_j^{-} \geq 0.
\]

\texttt{LassoPlus} is a Bayesian sparse modeling approach that simultaneously selects and estimates effects. The key feature of \texttt{LassoPlus} lies in its capability to estimate higher-order interaction terms without including the lower-ordered terms comprised of it. This property is particularly useful in subgroup analysis since researchers might be interested in the interaction terms between the predictors and the subgroup identifiers even when one of them in itself is not significant or of interest.

\begin{figure}[H]
  \centering
  \includegraphics[width= 0.9 \textwidth]{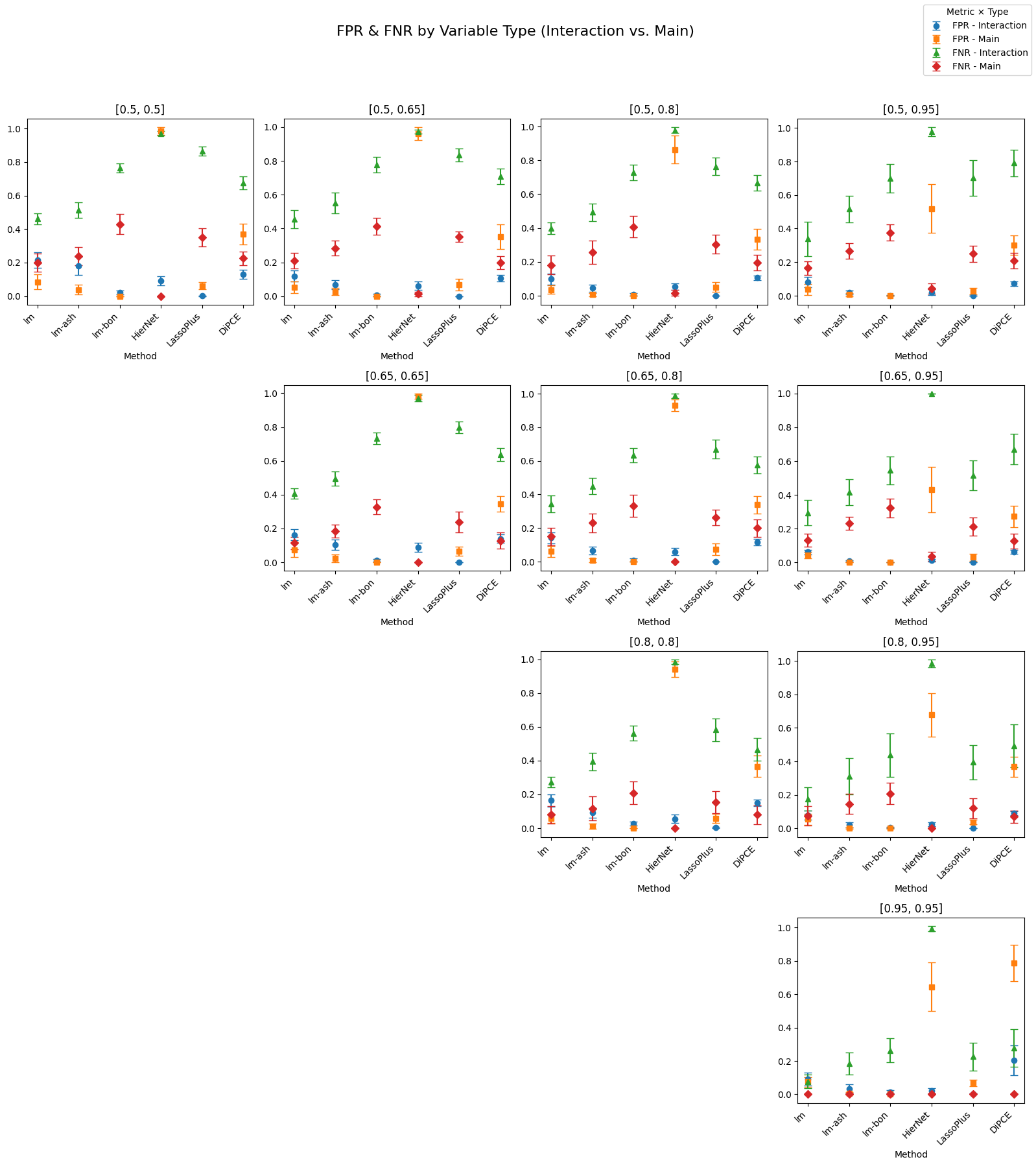}
  \caption{False positive rate (FPR) and false negative rate (FNR) by variable type (main vs. interaction) across all sparsity configurations $[Sp_j, Sp_z]$. Each row in the figure corresponds to a fixed $Sp_j$ level, and each column corresponds to a fixed $Sp_z$.}
  \label{fig:fpr_fnr}
\end{figure}

\section{Results}

\subsection{Coefficient Recovery Results }

\texttt{DiPCE} strikes an effective balance between false positives and false negatives for interaction effects under moderate sparsity assumptions about the degree of non-zero effects. In Figure~\ref{fig:fpr_fnr}, false positive rates (FPR) are shown in blue and false negative rates (FNR) in green for interaction effects. When interaction sparsity ($Sp_z$) is moderate (e.g., $Sp_z = 0.5$, $0.65$, $0.8$), \texttt{LassoPlus} consistently achieves the lowest FPR, followed by \texttt{lm-bon}, \texttt{DiPCE}, \texttt{HierNet}, \texttt{lm-ash}, and \texttt{lm}. The FNR rankings tell a different story: \texttt{HierNet} exhibits the highest FNR, indicating frequent failure to identify true interaction effects, followed by \texttt{LassoPlus}, \texttt{lm-bon}, \texttt{DiPCE}, \texttt{lm-ash}, and \texttt{lm}. Among all methods, \texttt{DiPCE} and \texttt{lm-bon} offer the most favorable trade-offs, achieving both low FPR and relatively low FNR. Notably, \texttt{DiPCE} is particularly effective at recovering true interaction effects while maintaining a modest false positive rate. However, no single method dominates across both metrics: improving FNR often comes at the cost of a higher FPR, and vice versa. Interestingly, \texttt{HierNet} reverses its pattern from the main effect case, it has a low FPR but very high FNR on interaction effects, suggesting it becomes overly conservative when detecting interactions.

Under extreme interaction sparsity, most methods, including \texttt{DiPCE}, effectively control false positives but struggle with false negatives. When $Sp_z = 0.95$, nearly all interaction effects are null, and most methods maintain low FPRs. The exception is \texttt{lm}, which continues to show slightly elevated false positives. FNRs increase for all methods due to the inherent difficulty of recovering very sparse signals. Confidence intervals also widen, reflecting the increased uncertainty in these sparse settings. This underscores the challenge of interaction detection when signals are weak and infrequent, even for methods like \texttt{DiPCE} that otherwise perform well in moderate settings.

For main effects, \texttt{DiPCE} trades off higher FPR for lower FNR, particularly under moderate sparsity. In Figure~\ref{fig:fpr_fnr}, false positive rates (FPR) are shown in orange and false negative rates (FNR) in red for main effects. When $Sp_j = 0.5$, \texttt{lm-bon} nearly eliminates false positives, achieving the lowest FPR, followed by \texttt{lm-ash}, \texttt{LassoPlus}, \texttt{lm}, \texttt{DiPCE}, and \texttt{HierNet}. However, in terms of FNR, \texttt{HierNet} performs best, followed by \texttt{DiPCE}, \texttt{lm}, \texttt{lm-ash}, \texttt{LassoPlus}, and finally \texttt{lm-bon}. These patterns underscore key trade-offs: \texttt{HierNet} is aggressive in flagging effects (low FNR, high FPR), while \texttt{lm-bon} is overly conservative (low FPR, high FNR). \texttt{lm-ash} occupies a middle ground, offering better balance. \texttt{DiPCE} is slightly less conservative in FPR than \texttt{lm-ash}, achieving low FNRs, recovering many true effects, but at the cost of more false positives.

\texttt{DiPCE}'s asymmetric performance across main and interaction effects reflects its dependence on accurately estimating interactions. While \texttt{DiPCE} performs well in recovering true interaction effects, its FPR for main effects is relatively higher. We hypothesize this is due to its estimation procedure: main effect recovery in \texttt{DiPCE} depends on accurate estimates of its associated interaction terms. Since other models estimate main effects directly, they do not suffer from this dependency. As sparsity increases (e.g., $Sp_j = 0.65$ and $0.8$), all models show rising FNRs. \texttt{lm-bon} remains the most conservative, with the lowest FPR and highest FNR. \texttt{HierNet}, conversely, maintains the lowest FNR but highest FPR, consistent with its aggressive behavior. In the high-sparsity regime ($Sp_j = 0.95$), where very few main effects are non-zero, all models correctly identify true positives with low FNR, and \texttt{lm-bon} and \texttt{lm-ash} still suppress FPRs effectively.

\subsection{Predictive Accuracy}
\label{sec:predicted_accuracy_results}

\begin{figure}[ht]
\centering
\includegraphics[width=1\linewidth]{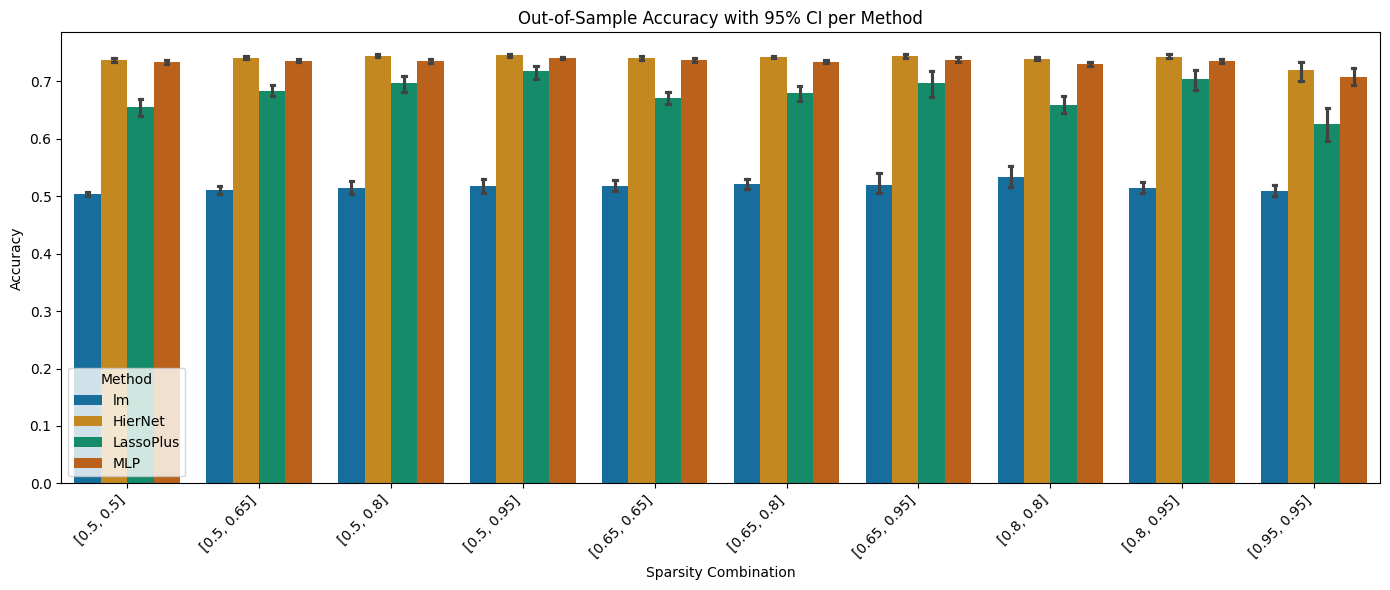}
\caption{Out-of-sample test accuracy across methods (\texttt{lm}, \texttt{HierNet}, \texttt{LassoPlus}, \texttt{MLP}) by sparsity configurations [$Sp_j$, $Sp_z$], where ${Sp_j}$ denotes the number of active main effects and ${Sp_z}$ denotes the number of active interaction effects.}
\label{fig:accuracy}
\end{figure}

Among all models, \texttt{DiPCE} and \texttt{HierNet} achieve the highest out-of-sample predictive accuracy, while linear regression lags due to its unsuitability for binary classification and its inability to learn nonlinear relationship. Figure~\ref{fig:accuracy} shows the out-of-sample accuracy for each method across different sparsity configurations. The bars represent the average test accuracy across 15 simulation replicates for each configuration, and the error bars indicate 95\% confidence intervals. The relatively poor classification accuracy of linear regression (\texttt{lm}) is expected, given that it is not designed for binary classification. We include it here because it remains the most commonly used estimator for coefficients in conjoint analysis. As shown in Figure~\ref{fig:accuracy}, both \texttt{HierNet} and \texttt{DiPCE} achieve strong predictive performance, consistently exceeding 70\% accuracy in most settings. Their performance is nearly identical in low-sparsity configurations, but \texttt{HierNet} slightly outperforms \texttt{DiPCE} as sparsity increases. \texttt{LassoPlus} performs consistently in the 60–70\% range, lower than \texttt{HierNet} and \texttt{DiPCE}, but still well above chance level (50\%) for a binary classification task.  It is worth noting that both \texttt{HierNet} and \texttt{LassoPlus} are explicitly provided with the information that interaction terms exist between attributes and context variables. In contrast, the \texttt{DiPCE} model is trained only on the raw one-hot encoded data and must implicitly learn any interaction effects on its own. 

\begin{figure}[t]
\centering
\includegraphics[width=1\linewidth]{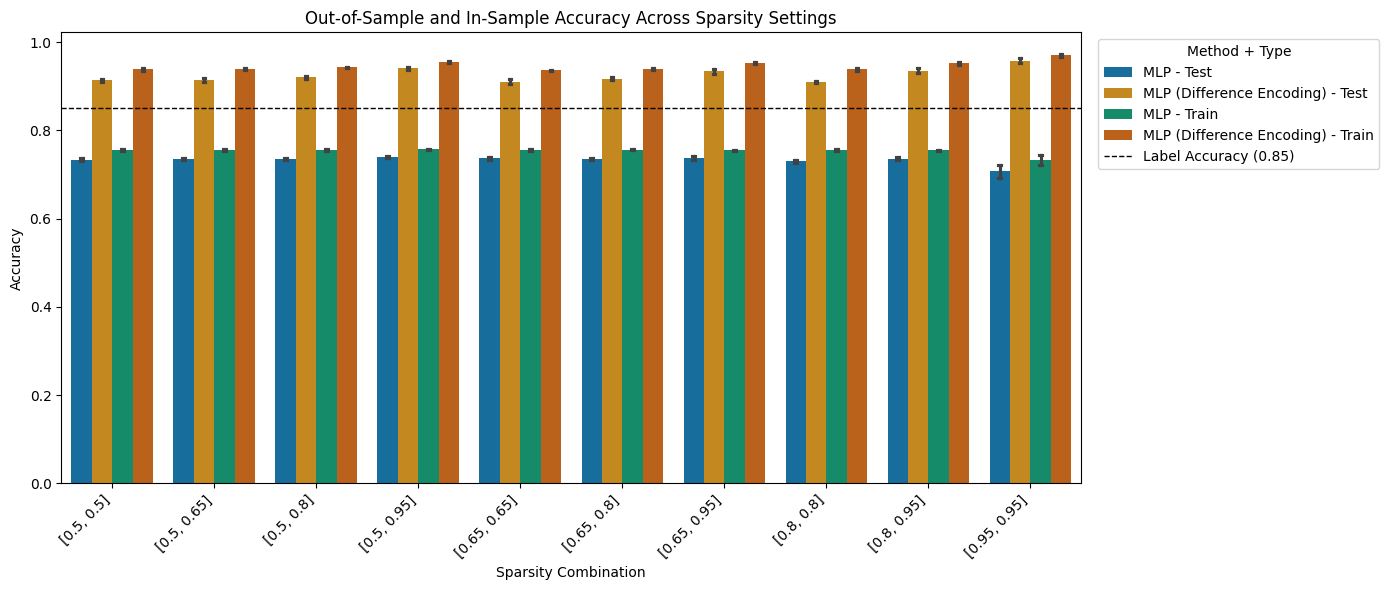}
\caption{Train and test accuracy of \texttt{DiPCE} models with and without difference encoding across sparsity configurations $[Sp_j, Sp_z]$.}
\label{fig:diff encoding}
\end{figure}

Finally, we find that difference encoding substantially boosts predictive accuracy, enabling \texttt{DiPCE} to surpass the theoretical label accuracy ceiling despite label noise.
Figure~\ref{fig:diff encoding} presents the training and test accuracy of MLP models trained with and without difference encoding. Although the tasks differ slightly, the difference-encoded model is trained on reduced representations of profile pairs, the comparison still reveals clear patterns. 
The black dotted line at 85\% represents the theoretical upper bound on label accuracy due to the presence of measurement error. The fact that both training and test accuracy exceed this 85\% threshold indicates that the models are able to recover true signal despite noisy labels. 

\section{Conclusion}
Conjoint analysis has become increasingly popular over the past decade among social and political researchers. As the methodology has evolved, more sophisticated experimental designs have emerged, most notably those that incorporate contextual or subgroup variables alongside traditional attribute profiles. However, to our knowledge, no existing work has introduced methods specifically tailored to this newer class of context-aware conjoint designs. Conveniently, our approach also addresses a problem that \emph{is} well-studied, namely subgroup analysis. Additionally, while it is plausible that existing approaches may be adapted to estimate heterogeneous effects in these settings, a systematic evaluation has been lacking. To address this gap, we develop an open-source simulation framework that provides flexible and realistic testbeds for benchmarking various estimation techniques across a wide range of sparsity configurations. Our simulations allow for a comprehensive comparison of several leading methods, including linear regression with $p$-value correction \citep{liu2023multiple}, Bayesian lasso \citep{ratkovic2017sparse}, and hierarchical lasso \citep{bien2013lasso}. Across multiple sparsity conditions, we find that each of these methods exhibits suboptimal performance in terms of either false positive rate (FPR) or false negative rate (FNR). These outcomes are not entirely surprising, as many of the approaches rely on assumptions about sparsity patterns or effect structures that align with their respective original applications, assumptions which only hold in specific subsets of our simulations. Notably, no existing method consistently provides robust performance across all design settings.

To overcome this limitation, we propose a predictive modeling-based causal inference method that leverages flexible neural networks to capture the complex, nonlinear structure of the data. Our approach recovers coefficient estimates for both main and interaction effects by carefully manipulating conditional expectations of predicted outcomes. While the method’s performance on main effects is limited, partly due to its dependence on accurate estimation of related interaction terms, it performs well in identifying significant interaction effects under moderate sparsity. It achieves low FPR and reasonable FNR in most settings, even in the presence of moderate label noise. We also introduce a difference encoding technique that exploits the pairwise structure of forced-choice conjoint tasks, enabling more accurate estimation by modeling profile comparisons directly. Finally, we believe that future research can build on our findings by combining the strengths of different methods, incorporating ideas from recent advances in double machine learning (DML) to reduce regularization bias, and leveraging the randomization inherent in conjoint designs to improve computational efficiency.
\section{Data Availability Statement}
Replication code for this article will be made available at the requisite Harvard Dataverse citation upon acceptance. All data used for this study was simulated and thus is replicable from the code only.

\section{Generative AI Disclosure}
Generative AI tools were used to assist in (i) searching for relevant research papers 
(alongside non-generative tools such as Google Scholar), (ii) debugging code, 
(iii) checking and correcting grammar and spelling, and (iv) \LaTeX{} formatting and spacing. 
All AI-assisted outputs were reviewed and manually edited by the authors.
All AI usage complies with Cambridge University Press guidelines on AI contributions to research.




\title{Appendices}

\maketitle
\section*{Appendix A: Evaluating Positive vs. Negative Encoding in MLP Estimators}
\label{sec:appendixB}

\begin{figure}[ht]
  \centering
  \includegraphics[width=\textwidth]{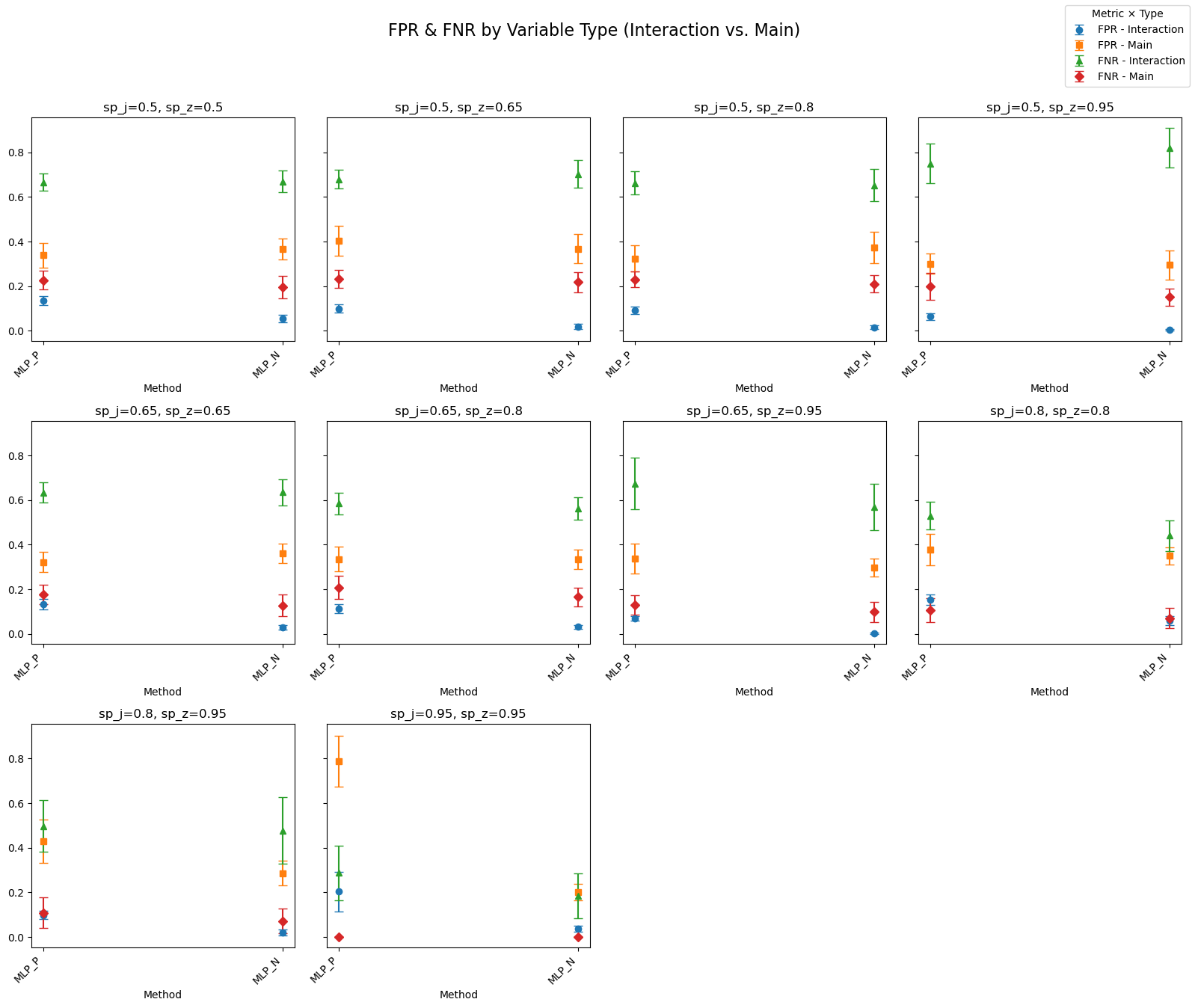}
  \caption{Comparison of false positive rates and false negative rates between two MLP-based estimators under difference encoding: MLP\_P, which uses $\mathbb{E}[Y \mid X_k^D = 1]$, and MLP\_N, which uses $-\mathbb{E}[Y \mid X_k^D = -1]$. }
  \label{fig:mlp_p_vs_n}
\end{figure}

Figure~\ref{fig:mlp_p_vs_n} presents a comparison between two variants of our proposed MLP-based estimator: MLP\_P and MLP\_N. The difference arises from how we compute the expected value $\mathbb{E}[Y \mid X_k^D = d]$ for difference-encoded variables, where $X_k^D \in {-1, 0, 1}$. Specifically, MLP\_P corresponds to using $\mathbb{E}[Y \mid X_k^D = 1]$, which represents the case where the left profile has $X_k = 1$ and the right profile has $X_k = 0$. Conversely, MLP\_N is based on $-\mathbb{E}[Y \mid X_k^D = -1]$, which corresponds to the reverse: left profile with $X_k = 0$ and right profile with $X_k = 1$.

In theory, assuming infinite data and a perfectly symmetric model, these two formulations should yield identical estimates. However, in practice, neural networks do not inherently encode this symmetry, and differences may emerge due to asymmetries in the training data or optimization process. Figure~\ref{fig:mlp_p_vs_n} shows that MLP\_N generally achieves lower false positive rates while maintaining comparable false negative rates across most sparsity configurations. This suggests it is a more conservative and reliable variant for inference. Based on this empirical evidence, we adopt MLP\_N as the default approach in our method.

\clearpage

\section*{Appendix B: Neural network used in this study}
\label{sec:appendixA}

\begin{algorithm}[H]
\caption{MLPClassifier}
\label{alg:mlp_classifier}
\KwIn{Input dimension $d$, learning rate $\eta$}
\KwOut{Trained MLP classifier}

Initialize model with 3-layer MLP:\;
\Indp
Layer 1: Linear($d \rightarrow 128$), BatchNorm, ReLU, Dropout(0.3)\;
Layer 2: Linear(128 $\rightarrow$ 64), BatchNorm, ReLU, Dropout(0.2)\;
Layer 3: Linear(64 $\rightarrow$ 1)\;
\Indm

Set loss function to BCEWithLogitsLoss\;
Set optimizer to Adam with learning rate $\eta$\;

\vspace{1mm}
\SetKwFunction{FForward}{Forward}
\SetKwProg{Fn}{Function}{:}{}
\Fn{\FForward{$x$}}{
    \Return model output logits for input $x$\;
}

\vspace{1mm}
\SetKwFunction{FTrain}{TrainingStep}
\Fn{\FTrain{$(x, y)$}}{
    logits $\leftarrow$ \FForward{$x$}\;
    loss $\leftarrow$ BCEWithLogitsLoss(logits, $y$)\;
    preds $\leftarrow$ sigmoid(logits) $> 0.5$\;
    acc $\leftarrow$ mean(preds == $y$)\;
    Log train\_loss and train\_acc\;
    \Return loss\;
}

\vspace{1mm}
\SetKwFunction{FOpt}{ConfigureOptimizers}
\Fn{\FOpt{}}{
    \Return Adam optimizer with learning rate $\eta$\;
}

\end{algorithm}

\section*{Appendix C: Estimation Details for DiPCE}~\label{sec:appendix_dipce}

\subsection*{Adjusted Derivation}
With the introduction of difference encoding, the estimation procedure outlined in Section~\ref{sec:derivation} must be adjusted accordingly. Specifically, all instances of $X_k$ must be replaced with the corresponding difference-encoded variable $X_k^D$. For simplicity, we illustrate this using the core expected value difference from Section~\ref{sec:derivation}, namely Equation~\ref{eq:xk1-xk0}, which computes the quantity:
\[
\mathbb{E}[Y \mid X_k = 1] - \mathbb{E}[Y \mid X_k = 0].
\]
Under difference encoding, this comparison is reformulated in terms of $X_k^D$. In particular, we evaluate $\mathbb{E}[Y \mid X_k^D = 1]$,which corresponds to cases where the left profile has $X_k = 1$ and the right profile has $X_k = 0$, and $\mathbb{E}[Y \mid X_k^D = -1]$, where the roles are reversed. Because these two conditions represent mirror-image comparisons, we compute the adjusted quantity as:
\[
\mathbb{E}[Y \mid X_k^D = 1] \text{ and }  (-1) \cdot \mathbb{E}[Y \mid X_k^D = -1],
\]

It is important to note that we exclude observations where $X_k^D = 0$, which reflects cases where both profiles share the same level of $X_k$ and thus provide no information for estimating its effect. While this exclusion slightly departs from the original derivation, the substantive quantity being estimated remains the same: the contrast in outcomes between profiles with $X_k = 1$ and those with $X_k = 0$. By conditioning only on informative comparisons (i.e., those where $X_k^D \neq 0$), we improve efficiency without altering the interpretation of the estimated effect.

\subsection*{Estimation}

To determine whether a coefficient is statistically significant, we estimate its value along with a corresponding confidence interval. To construct confidence intervals, we apply a bootstrap procedure to the quantities involved in estimating main and interaction effects. Specifically, we resample the data to estimate quantities such as $\mathbb{E}[Y \mid X_k^D = 1]$, $\mathbb{E}[Y \mid X_k^D = -1]$, $\mathbb{E}[Y \mid X_k^D = 1, C_l = 0]$, and $\mathbb{E}[Y \mid X_k^D = -1, C_l = 0]$.

The bootstrap procedure is conceptually straightforward. Consider the estimation of the interaction coefficient $B_{k,l}$ between attribute $X_k$ and context variable $C_l$. As described in Section~\ref{sec:derivation}, this coefficient is computed using Equation~\ref{eq:two_exp_sub}:
\[
B_{k,l} = \left( \mathbb{E}[Y \mid X_k = 1] - \mathbb{E}[Y \mid X_k = 0] \right) - \left( \mathbb{E}[Y \mid X_k = 1, C_l = 0] - \mathbb{E}[Y \mid X_k = 0, C_l = 0] \right).
\]

To estimate this quantity, we first subset the test dataset to include only observations where $X_k^D = 1$ or $X_k^D = -1$ (refer to Appendix~\ref{sec:appendixA} in the supplementary materials), and obtain the predicted probabilities from the model by applying the sigmoid function to the logits. We subtract 0.5 from these predicted probabilities to center them around zero, this recentering makes it easier to interpret the magnitude and direction of each variable’s influence on the probability that a profile is chosen. Using predicted probabilities instead of discrete labels is a common approach in causal inference, as it provides higher resolution information about effect magnitudes. This strategy has been employed in studies estimating the causal effects of binary exposures \citep{le2021g} and group marginal effects \citep{long2021using} for binary outcomes .

After adjusting the predicted probabilities, we draw bootstrap samples and compute $B_{k,l}$ for each resampled dataset using Equation~\ref{eq:two_exp_sub}. Repeating this process over many resamples yields a distribution of coefficient estimates, from which we construct percentile-based confidence intervals. If the resulting confidence interval contains zero, we conclude that the coefficient is not statistically significant; otherwise, we consider the effect to be significant. We evaluated several machine learning models, including random forest, XGBoost, CatBoost, and a multilayer perceptron (MLP). We selected the MLP as it achieved the highest predictive accuracy. A detailed outline of our estimation procedure is provided in Algorithm~\ref{alg:bootstrap_effects}.

\begin{algorithm}

\caption{Bootstrapped Estimation of Main and Interaction Effects with 95\% CI Filtering}
\label{alg:bootstrap_effects}
\KwIn{
Model $f$, main effect list $\mathcal{M}$, context list $\mathcal{C}$, column names $\mathcal{X}$, observation matrix $X$, number of bootstraps $B$
}
\KwOut{
Filtered estimates where CI$_{95\%} \not\ni 0$
}

\textbf{Step 1: Predict Centered Logits}\;
Initialize $D \gets \emptyset$\;
\ForEach{$m \in \mathcal{M}$}{
  \ForEach{$c \in \mathcal{C}$}{
    Find column indices $i_m, i_c$ in $\mathcal{X}$\;
    Define mask $M = \{x \in X \mid x_{i_m}=1 \land x_{i_c}=1\}$\;
    Compute logits $Y = \sigma(f(X[M])) - 0.5$\;
    Add rows $(Y, m, c)$ to $D$\;
  }
}

\vspace{0.5em}
\textbf{Step 2: Bootstrap Estimation}\;
Let $p \gets \frac{1}{|\mathcal{C}|}$, and initialize result set $R \gets \emptyset$\;

\ForEach{$m \in \mathcal{M}$}{
  Let $D_m \gets \{d \in D \mid d.\text{Main} = m\}$\;
  Initialize $\Delta \gets [\phantom{0}]$, and $\Delta_{m:c} \gets [\phantom{0}]$ for all $c \in \mathcal{C}$\;

  \For{$b = 1$ \KwTo $B$}{
    Sample $D_b \sim D_m$ with replacement\;
    Let $\bar{Y}_A \gets \text{mean}(D_b.Y)$\;

    \ForEach{$c \in \mathcal{C}$}{
      Let $B_i \gets \{y \in D_b.Y \mid D_b.\text{Context} \neq c\}$\;
      If $|B_i| > 0$, compute:
      $\bar{Y}_B^{(c)} = \text{mean}(B_i)$\;
      Append $\frac{\bar{Y}_A - \bar{Y}_B^{(c)}}{p}$ to $\Delta_{m:c}$\;
    }

    If all $\bar{Y}_B^{(c)}$ exist, compute:
    $Z^\ast = |\mathcal{C}| \cdot \bar{Y}_A - \sum_{c} \bar{Y}_B^{(c)}$\;
    Append $\bar{Y}_A - Z^\ast$ to $\Delta$\;
  }

  \ForEach{effect $\theta \in \{\Delta\} \cup \{\Delta_{m:c} \;|\; c \in \mathcal{C}\}$}{
    Compute 95\% CI: $[\theta_{2.5}, \theta_{97.5}]$ and mean $\bar{\theta}$\;
    \If{$0 \notin [\theta_{2.5}, \theta_{97.5}]$}{
      Add $(\bar{\theta}, \text{CI}_{95\%})$ to $R$\;
    }
  }
}

\Return{$R$}

\end{algorithm}

\end{document}